\documentclass[%
reprint,
 amsmath,amssymb,
 aps,
prb,
]{revtex4-2}

\usepackage{bm}
\usepackage{graphicx}% Include figure files
\usepackage{dcolumn}% Align table columns on decimal point
\usepackage{bm}% bold math
\usepackage{hyperref}
\usepackage{subfigure}

\begin{document}

\title{ScGAN: A Generative Adversarial Network to Predict Hypothetical Superconductors}
\date{\today}

\author{Evan Kim}
    \email{evan.e.kim@gmail.com}
\affiliation{Tesla STEM High School, Redmond, WA 98053, USA}
 
 \author{S.V. Dordevic}
    \email{dsasa@uakron.edu}
\affiliation{Department of Physics, The University of Akron, Akron, OH 44325, USA}

\begin{abstract}
Despite having been discovered more than three decades ago, High Temperature Superconductors (HTSs) lack both an explanation for their mechanisms and a systematic way to search for them. To aid this search, this project proposes ScGAN, a Generative Adversarial Network (GAN) to efficiently predict new superconductors. ScGAN was trained on compounds in OQMD and then transfer learned onto the SuperCon database or a subset of it. Once trained, the GAN was used to predict superconducting candidates, and approximately 70\% of them were determined to be superconducting by a classification model--a 23-fold increase in discovery rate compared to manual search methods. Furthermore, more than 99\% of predictions were novel materials, demonstrating that ScGAN was able to potentially predict completely new superconductors, including several promising HTS candidates. This project presents a novel, efficient way to search for new superconductors, which may be used in technological applications or provide insight into the unsolved problem of high temperature superconductivity.
\end{abstract}

\maketitle

\section{Introduction}
\label{introduction}

In recent years, superconductors have been applied in a variety of important technologies such as power transmission lines, MRI magnets, Maglev trains, and quantum computers. Quantum computers are especially important as they are expected to solve problems that are too computationally expensive for current classical computers. However, to be used in these applications the superconductors must be cooled below their critical temperatures ($T_c$), which for most current superconductors are very low. For instance, Google's quantum computer must be maintained at $0.02 \;\mathrm{K}$, which severely limits its general use \cite{Arute2019}. This points to a growing need for superconductors with higher $T_c$, 
which has made them an active research topic for the last couple of decades.

The mechanisms of superconductivity in most materials with relatively high $T_c$ is not fully understood, which means there is no systematic way to search for new materials or to predict their critical temperatures \cite{HIRSCH20151}. Thus, the current procedure for finding HTSs is essentially trial--and--error, which is extremely inefficient. This was exemplified in a recent study, which found that only about $3\%$ of the approximately $1000$ materials surveyed were superconducting \cite{Hosono2015}. Furthermore, the study failed to find any superconductors with $T_c>60 \; \mathrm{K}$. This extreme inefficiency means that the likelihood of manually finding new superconductors, especially HTSs, is extremely low. 

To address this difficulty, the use of computational tools in superconductivity research has become popular in recent years \cite{Bedolla_2020}. In particular, there have been several studies utilizing machine learning to predict whether a given chemical compound will be superconducting or not. The earliest such study was done by \citeauthor{Stanev2018} in which two random forest-based models were built: a classification model for predicting superconductivity and a regression model for predicting superconducting transition temperature \cite{Stanev2018}. The models were successfully applied on the SuperCon database, achieving a 92\% accuracy in classification and a $R^2=0.88$ for regression. They did run across one limitation of their machine learning model, however. When trained on a certain class of superconductors (e.g. cuprates), the model was unable to make good predictions on other classes of superconductors (e.g. pnictides).

Following this pioneering study, there have been several other studies applying machine learning methods, such as a K-nearest neighbors algorithm (Ref.~\cite{ROTER20201353689}) and a deep learning model (Ref.~\cite{PhysRevB.103.014509}). The K-nearest neighbors algorithm reported improvements on the previous study from Ref.~\cite{Stanev2018} in terms of overall performance: an $R^2$ of $0.93$ and a classification accuracy of $96.5\%$. The deep learning model, on the other hand, showed that it might be possible to overcome the limitation that \citeauthor{Stanev2018} faced, as they were able to make predictions about pnictide superconductors from the training data that did not contain them.

The way each of these previous studies predicted new superconducting materials was by running their trained model on a database of known existing chemical compounds and finding which compounds the model indicated could be superconducting. This procedure has several notable limitations. First, these studies miss out on the vast chemical composition space that is not already contained in existing databases. Second, commonly used databases (such as ICSD and OQMD) contain mostly stoichiometric compounds, whereas many superconductors are non-stoichiometric (cuprates and pnictides, in particular), and so many possibilities were missed in that manner as well (see for example Table 3 in Ref.~\cite{Stanev2018} and Table 1 in Ref.~\cite{ROTER20201353689}). Finally, the discovery of superconductors in the lab usually does not happen that way; they are discovered by synthesizing new materials and testing them, rather than checking the known ones.

In order to overcome these limitations, in this work we employ Generative Adversarial Networks (GANs) \cite{goodfellow2014}. Generative models refer to a general class of machine learning models which are able to generate things that resemble the input dataset. They have proven to be extremely powerful, and in recent years have found numerous applications such as science, engineering, medicine, art, video games, deepfakes, etc. \cite{DBLP:ganreview}. In most cases GANs performed better than other generative models, such as variational autoencoders, because they are able to learn more hidden rules in the input data set. For example, \citeauthor{Dan2020} \cite{Dan2020} reported a GAN model which generated new chemical compounds with 92.53\% novelty and 84.5\% validity. Another work, \citeauthor{sym12111889} \cite{sym12111889} applied GANs to the SuperCon dataset, but for the purpose of characterization rather than prediction. 

In this work, we combine the general idea of GANs with the previous superconductor models, and propose the first GAN to predict new superconductors (ScGAN). Our ScGANs are based on chemical composition only, and are able to generate new superconducting materials with high novelty and validity. The paper is organized as follows. In Section \ref{methodology} we present the details of the creation of the ScGANs. In section \ref{results} the main results of the study are discussed. In particular, we present a list of hypothetical superconducting materials generated by our ScGANs, as well as their predicted critical temperatures. Finally, in Section \ref{conclusions} we summarize the most important findings.

\section{Methodology}
\label{methodology}

\begin{figure*}
        \centering
        \includegraphics[scale=0.6]{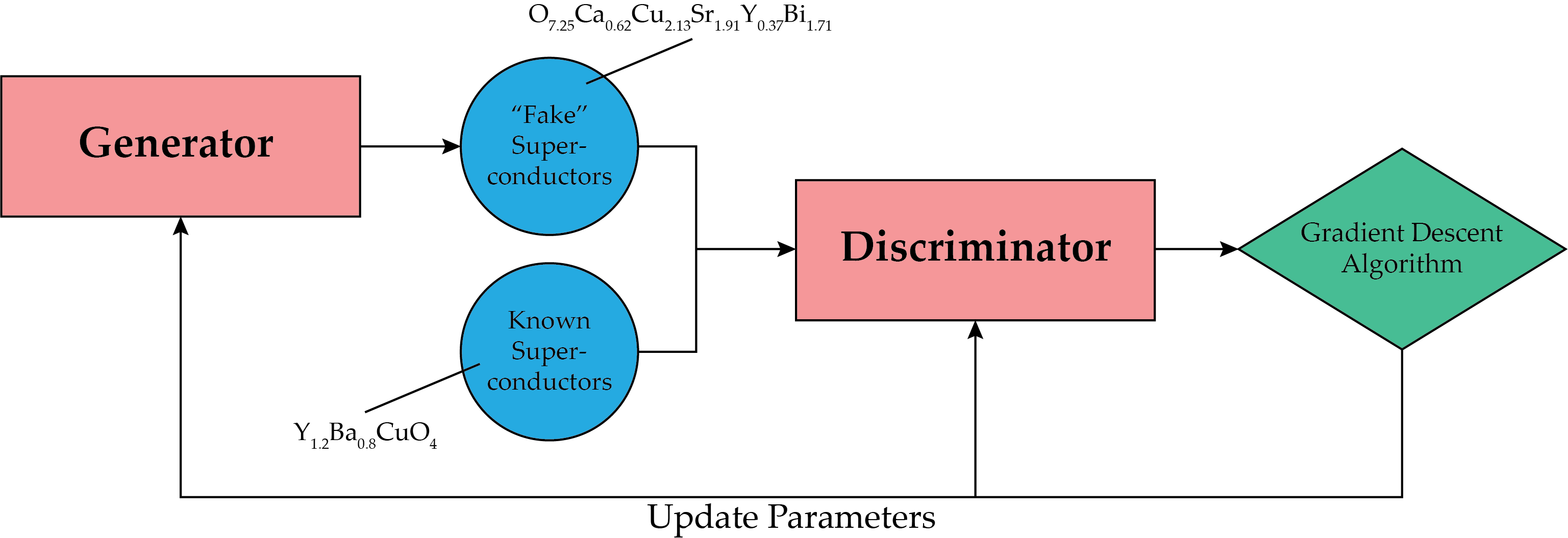}
        \caption{The architecture of the GAN model used. Y$_{1.2}$Ba$_{0.8}$CuO$_4$ and O$_{7.25}$Ca$_{0.62}$Cu$_{2.13}$Sr$_{1.91}$Y$_{0.37}$Bi$_{1.71}$ 
        are examples of real (from SuperCon) and ``fake'' superconductors, respectively. }
        \label{fig:ganarch}
    \end{figure*}

As stated in the introduction, we chose the GAN as our generative model. Its structure is shown in Fig. \ref{fig:ganarch}. A GAN is composed of two competing neural networks, the generator and discriminator. The generator takes in random noise and generates a ``fake'' compound, which the discriminator attempts to determine if it is real or not using existing data of real compounds. Each of them then updates their parameters based on the performance of the discriminator. The two networks improve their performance iteratively until a generator can generate realistic looking compounds, while the discriminator can detect unrealistic looking ones \cite{goodfellow2014}.

\begin{figure}
    \centering
    \includegraphics[scale=0.4]{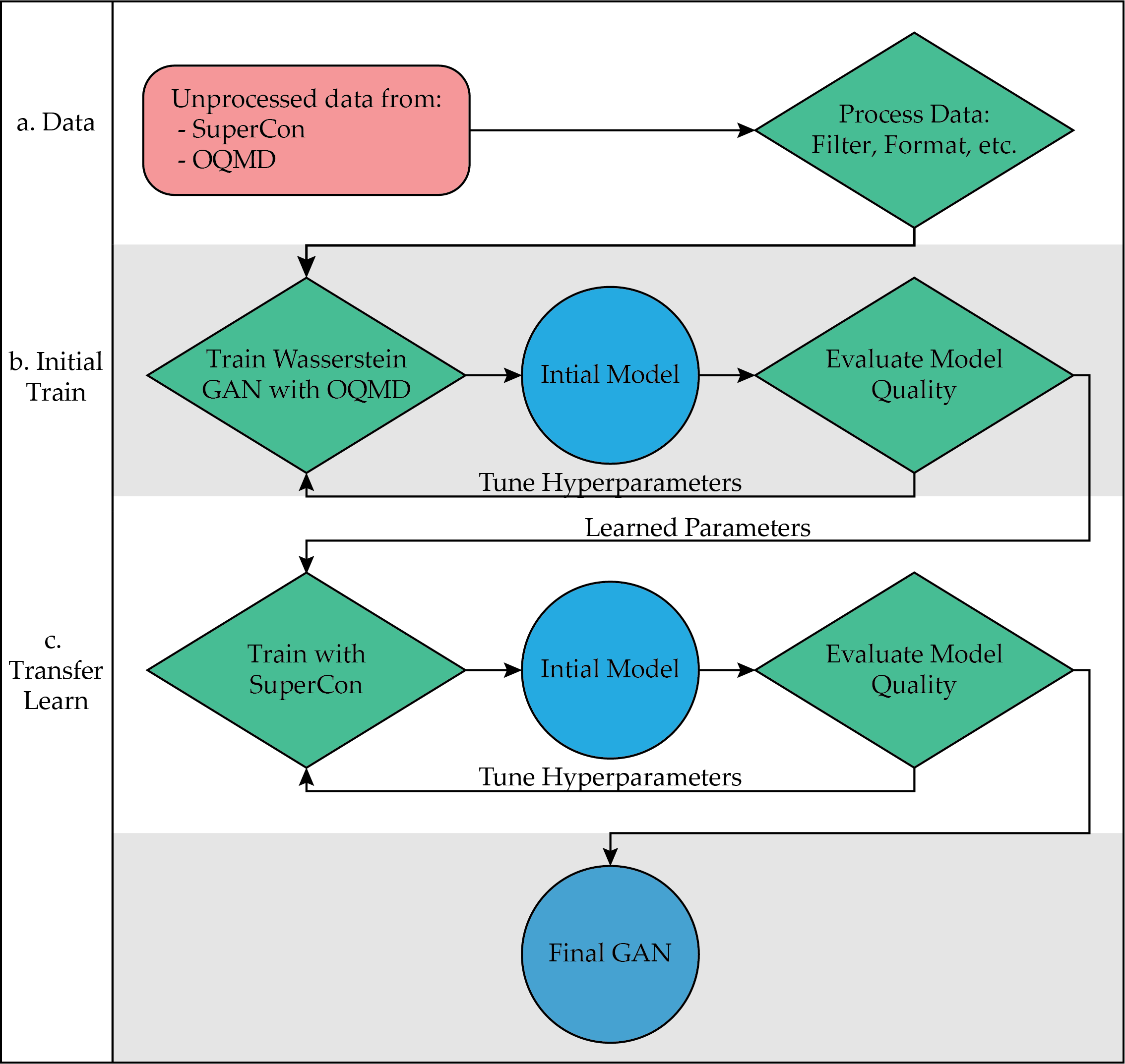}
    \caption{The overall training / testing process from the data to the final GAN model. It is composed of three main stages: (a) data processing, (b) training on the OQMD dataset, and finally (c) transfer learning onto the superconductor dataset (the whole SuperCon, or a part of it).}
    \label{fig:trainflowchart}
\end{figure}

Fig. \ref{fig:trainflowchart} depicts the training/testing process of our GANs and it has three main stages. The main idea is to have the model first learn general chemical composition rules by training off of a larger dataset (OQMD database) of general chemical compounds and then transfer learning it onto a (much smaller) dataset of known superconducting materials (SuperCon). The idea of transfer learning is to allow a model to learn even with a limited amount of data, which is the case here as the general compounds dataset (OQMD) has on the order of $\sim 10^6$ data points, while the superconductor datasets will only have on the order of $\sim 10^4$ data points.

\subsection{Data Collection}

Data were sourced from two open-source databases: SuperCon \cite{supercon.nims.go.jp} and the Open Quantum Materials Database (OQMD) \cite{Saal2013,Kirklin2015}. SuperCon is the largest database for superconducting materials with around $30,000$ superconductors before filtering. Similar to what was done in some previous studies \cite{ROTER20201353689,ROTER20221354078}, 
% \textbf{can we call it ``our previous projects'' even if I (evan) didn't participate in them?}, 
we only used the chemical compositions of the materials extracted from SuperCon. On the other hand, OQMD is a much larger database with around $10^6$ DFT-calculated compounds, most of which are not superconductors.

\subsection{Data Processing}
\label{sec:dataprocess}
\begin{figure}
    \centering
    \includegraphics[scale=0.28]{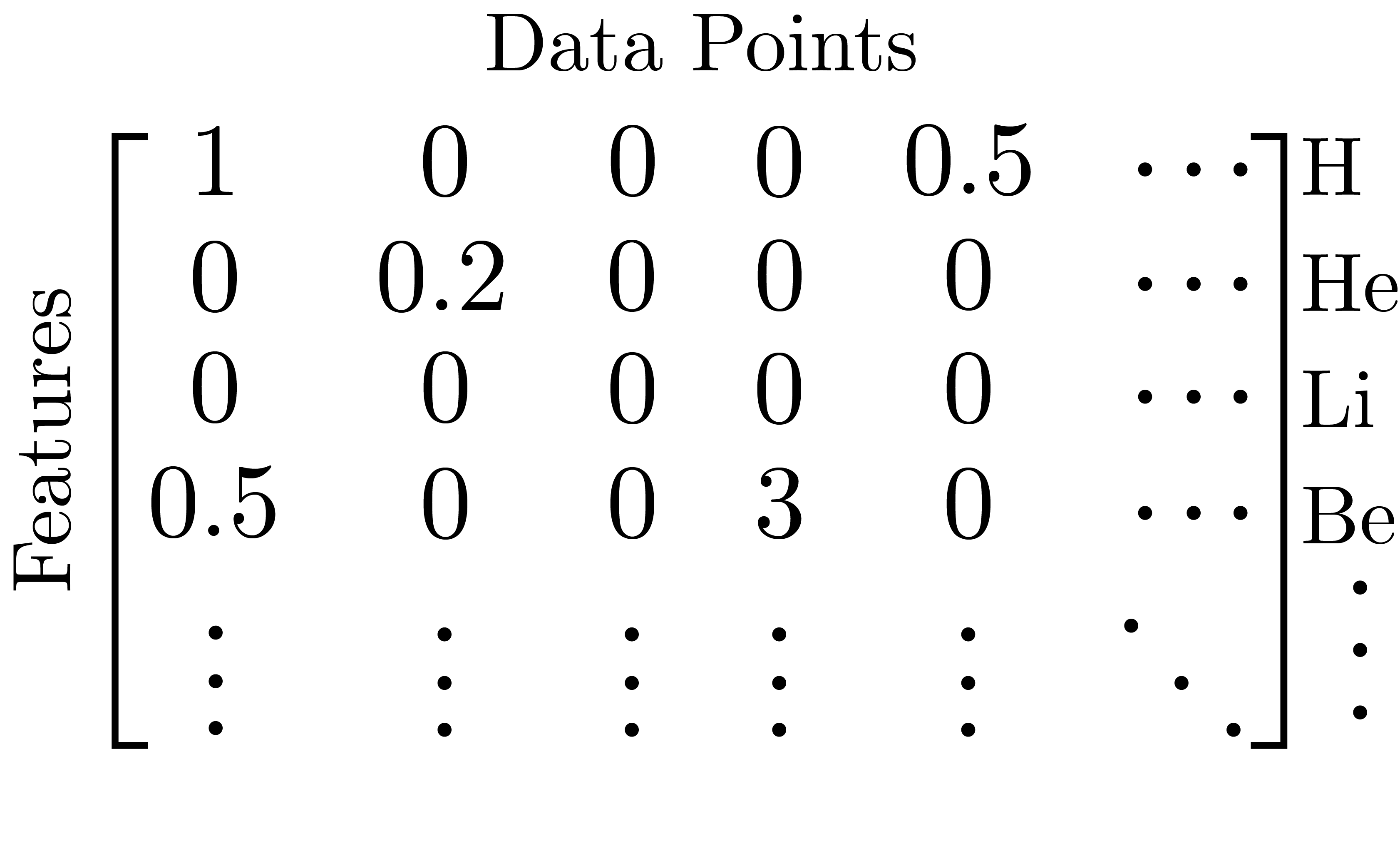}
    \caption{Chemical composition data represented as a matrix \cite{ROTER20201353689} in $\mathbb{R}^{96\times m}$, where $m$ is the number of datapoints. Each column is a single compound, with each entry representing the number of each element present in the compound. Note that that the numbers in the matrix are for illustration purposes only; they do not represent any real compounds or superconductors.}
    \label{fig:matrixrepresentation}
\end{figure}

\begin{figure}
    \centering
    \includegraphics[scale=1]{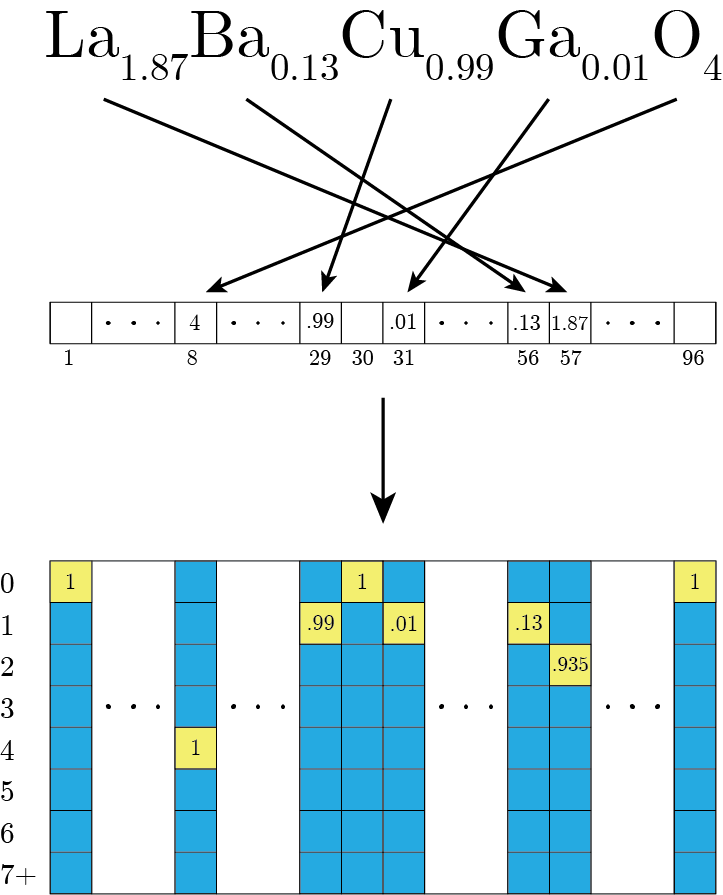}
    \caption{An ``adjusted one-hot'' encoding of the chemical composition as a matrix in $\mathbb R ^{96 \times 8}$. The idea here is that the amount of each element is encoded through both the vertical location of the yellow box and the value in it, which allows the GAN to learn the chemical compositions better.}
    \label{fig:dataf}
\end{figure}

\begin{table}
    \centering
    \begin{ruledtabular}
    \begin{tabular}{ccc}
        Class & Quantity & Percentage  \\
        \hline
        Cuprates & 7,304 & 44.4\% \\
        Pnictides & 1,436 & 8.7\% \\
        Others & 7,749 & 47.0\% \\\hline
        Everything & 16489 & 100\%
    \end{tabular}
    \end{ruledtabular}
    \caption{Distribution of the superconductors in our filtered version of SuperCon by class. Each of these served as different training sets.
    We notice that the number of pnictide entries is much smaller compared with cuprates and others. }
    \label{tab:class_distributions}
\end{table}

Before filtering the data, all compounds had to be expressed in a common format so that unwanted datapoints could be detected regardless of small differences in formatting across the databases, such as the ordering of elements. First, all datapoints were formatted as $1 \times 96$ matrices
\cite{ROTER20201353689} (the $96$ is the maximum atomic number present across all the compounds in the database) and then combined so that each dataset was a matrix in $\mathbb{R}^{96 \times m}$, where $m$ is the number of data points (see Fig.~\ref{fig:matrixrepresentation}). Then, both datasets were filtered for duplicates, which reduced OQMD to around $800,000$ datapoints. The SuperCon dataset required further processing as a number of entries were incomplete, and they were either corrected or removed, leaving around 16,000 datapoints. Lastly, the SuperCon dataset was split into three different groups (classes): cuprates, pnictides (iron-based) and others (anything else that is neither a curpate nor a pnictide). Table \ref{tab:class_distributions} has the quantitative counts of these different groups. The purpose of this was to test the model's ability to learn the different classes of superconductors, especially the high temperature classes of cuprates and pnictides.

Once the data was filtered, the chemical compositions were then transformed into a form for the GAN to train on. A previous study, Ref.~\cite{Dan2020}, used a one-hot encoding for general chemical compounds. However, that encoding was designed for stoichiometric compounds, i.e. for integer values of parameters, 
and many superconductors are non-stoichiometric, i.e. have decimal compositions. Instead, we propose the use of a ``adjusted one-hot'' encoding that works for decimals,
in which each compound is represented by a 96 $\times$ 8 matrix of real numbers (Fig.~\ref{fig:dataf}). As shown in the figure, from the $1 \times 96$ vectors in the columns of 
Fig.~\ref{fig:matrixrepresentation}, each nonzero component of that vector $\mathbf{v}_i$ was expanded into an 8-dimensional vector with the following process. First the nearest integer $k$ between $1$ and $7$ inclusive to $\mathbf{v}_i$ was found. Then, the matrix values were set as 
\begin{equation}
    A_{mi} = \delta_{mk} \cdot \mathbf{v}_i/k,
\end{equation}
where $\delta$ here is the Kronecker delta, we index from $0$ (top left is $A_{00}$), and $m$ ranges from $0$ to $7$. For zero components ($\mathbf{v}_i = 0$), a $1$ was simply placed at $A_{0i}$. We point out that we tested other encodings, such as the one from Ref.~\cite{PhysRevB.103.014509}, but these were susceptible to mode collapse. The encoding proposed in this work successfully encodes decimal values both through the actual matrix entry and its location.

\subsection{Model}

A GAN is a type of generative model that has two competing neural networks, a generator and a discriminator, as shown in Fig.~\ref{fig:ganarch}. Traditional GANs, however, can suffer from issues such as mode collapse and gradient vanishing, so the Wasserstein GAN with Gradient Penalty \cite{DBLP:journals/corr/GulrajaniAADC17} was used instead. In the Wasserstein GAN with gradient penalty, the loss functions are
\begin{align}
    \mathrm{Loss}_D &= \underset{\boldsymbol{\tilde{x}} \sim \mathbb P_g}{\mathbb E} [D ( \boldsymbol{\tilde{x}})] - \underset{\boldsymbol{x} \sim \mathbb P_r}{\mathbb E} [D(\boldsymbol{x})]\\ &\quad + \lambda \underset{\boldsymbol{\hat{x}} \sim \mathbb P_{\boldsymbol{\hat x}}}{\mathbb E} [(\| \nabla_{\boldsymbol{\hat x}} D(\hat{\boldsymbol x}) \| - 1)^2], \nonumber \\
    \mathrm{Loss}_G &= - \underset{\boldsymbol{\tilde{x}} \sim \mathbb P_g}{\mathbb E} [D ( \boldsymbol{\tilde{x}})].
\end{align}
Here $D(x)$ represents the output of the discriminator and $\mathbb{E}$ is the expectation value (average). Then the parameters are updated with an optimizer,
\begin{align}
            w &\leftarrow w + \alpha  \cdot \operatorname{Optimizer}(w,\nabla_w \mathrm{Loss}_D),\\
            \theta &\leftarrow \theta - \alpha \cdot \operatorname{Optimizer}(\theta,\nabla_\theta \mathrm{Loss}_G),
\end{align}
where $w$ are the discriminator's parameters, $\theta$ is the generator's parameters, and $\alpha$ is the learning rate. RMSProp was chosen as the optimizer \cite{hintonrmsprop}, after testing out several other options, such as Adam \cite{kingma2014adam}.

\subsubsection{Training}
\begin{figure*}
    \centering
    \includegraphics[scale=0.8]{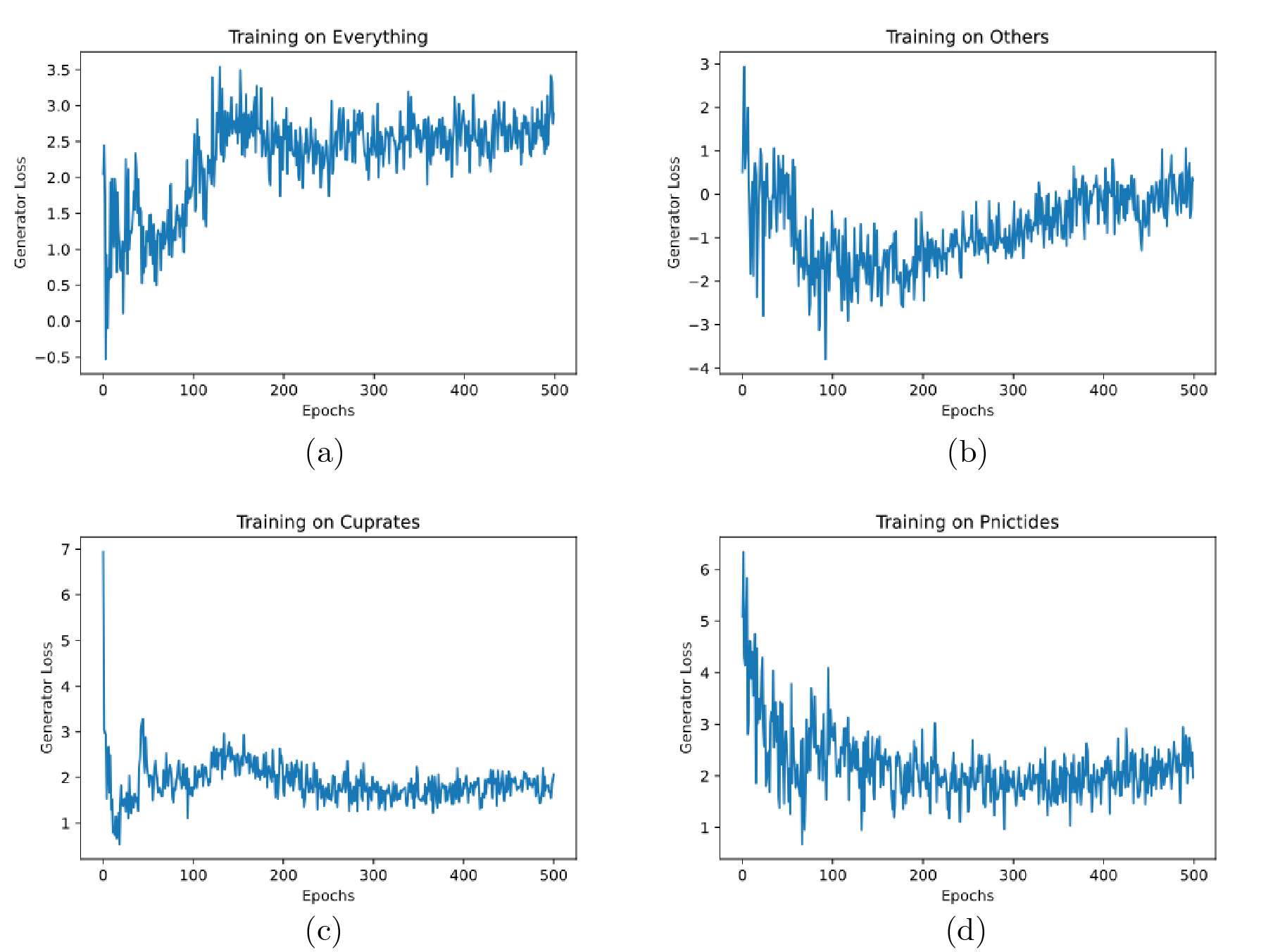}
    \caption{The generator loss against training epoch for each of the four datasets the GAN trained on: (a) All of SuperCon; (b) Others (i.e. not cuprates or pnictides); (c) Cuprates; and (d) Pnictides.}
    \label{fig:training_curves}
\end{figure*}

The model was first trained on the OQMD dataset for 400 epochs. It was then transfer learned onto the SuperCon dataset or a subset of it, on which it would train for another 500 epochs. Transfer learning onto four different datasets (cuprates, pnictides, others, and everything together) resulted in four different versions of the GAN. Afterwards, the testing procedure and the data analysis were conducted, and then the hyperparameters were updated based on the results. The training curves for the final model on each of these sets are displayed in Fig.~\ref{fig:training_curves}. Notably, they were all able to converge and stabilize over the 500 epochs.

\subsubsection{Testing}

After each training process, $5,000$ hypothetical compounds were generated from the versions of the GAN trained on the smaller sets and $30,000$ from the version trained on everything. The generated predictions were then inspected with various quality checks. Each compound was first tested for validity using the charge neutrality and electronegativity check features of the \texttt{SMACT} package \cite{DAVIES2016617}. The package tests the compound for electronegativity and charge balance to determine whether it is a valid compound or not. Each prediction was then checked for uniqueness---whether the compound showed up earlier in the generated list---and novelty---whether the compound was in the training dataset. These three checks looked at the general quality of the model. Then, more specific to superconductivity, each compound was run through the model from Ref.~\cite{ROTER20201353689} to check whether it is a superconductor or not, as well as to 
predict its critical temperature. Of course, to be sure of superconductivity, the compounds must be synthesized and tested. Lastly, the formation energy of each generated compound was calculated using the model from Ref.~\cite{Jha2018}, which indicates the stability of the compound. These tests will be discussed further in Section \ref{results}, along with the actual results.

\subsection{Clustering}

In order to further assess the quality of predictions,
clustering analysis was performed. Clustering is an unsupervised machine learning 
technique whose main goal is to unveil hidden patterns in the data.
It was recently applied to superconductors from SuperCon database \cite{ROTER20221354078}.
Depending on the data set, different clustering algorithms were used, such as k-means, hierarchical, Gaussian mixtures, 
self-organizing maps, etc. The results showed that in case of superconductors, clustering methods can achieve, 
and in some cases exceed, human level performance. 

In order to visualize clustering results, different techniques can be used. It was shown that for 
superconductors the so-called t-SNE produces the best results \cite{ROTER20221354078}.
t-SNE is a non-linear dimensionality reduction
technique which allows higher dimensional data to be represented in 2D or 3D \cite{vanDerMaaten2008}. 
In case of superconductors, the data points are $96$-dimensional (each compound is represented by a 1 $\times$ 96 matrix), 
as discussed in Section \ref{sec:dataprocess}. t-SNE reduces these dimensions down to either two or three, which allows easy visualization.  
We point out, however, that these reduced dimensions do not have any physical meaning.

\section{Results}
\label{results}

After training, from the four different versions of the GAN, we generated four superconductor candidate lists with either 5,000 or 30,000 chemical compositions. We then ran the predicted superconductors through a series of tests to evaluate their quality. The first few were general tests, and the rest were in the context of superconductivity using existing computational models, as experimentally testing all of them is unfeasible.

\subsection{Duplicates and Novelty}

We first screened the output for duplicates within the generated sets themselves and then for duplicates between the generated set and the dataset of known superconductors that it trained on. The results are tabulated in Table~\ref{tab:novelty}. As seen in the table, the number of repeats within the generated samples were relatively low (high uniqueness), with the exception of pnictides, which had more duplicates than the rest. This is likely due to the fact that the dataset of pnictides was significantly smaller than the rest (see Table~\ref{tab:class_distributions}). We speculate that this overall low rate of duplicates stems from the fact that the model is able to handle decimals (see Section IIB and Fig.4), which opens up a large composition space for it to explore.

The percent of predicted compounds that were novel, i.e. not already known to exist, listed in Table~\ref{tab:novelty} is also very high across all four GANs. These two results demonstrate that all versions of ScGAN can generate a \textit{diverse} array of \textit{new} compositions.

\begin{table}
    \centering
    \begin{ruledtabular}
    \begin{tabular}{ccc}
        GAN Version & Novel \% & Unique \%\\
        \hline
        Entirety of SuperCon & 99.69\% & 96.78\% \\
        Cuprate  & 99.74\% &  92.98\%\\
        Pnictides  & 99.32\% & 58.74\% \\
        Others &  98.89\% & 91.58 \%
    \end{tabular}
    \end{ruledtabular}
    
    \caption{The percentage of generated predictions that were novel (not in the training set) and  unique (distinct from others in the given generated set) for each of the versions of ScGAN trained on the given training datasets on the left.}
    \label{tab:novelty}
\end{table}

\subsection{Formation Energy}

\begin{figure*}
    \centering
    \includegraphics[scale=0.8]{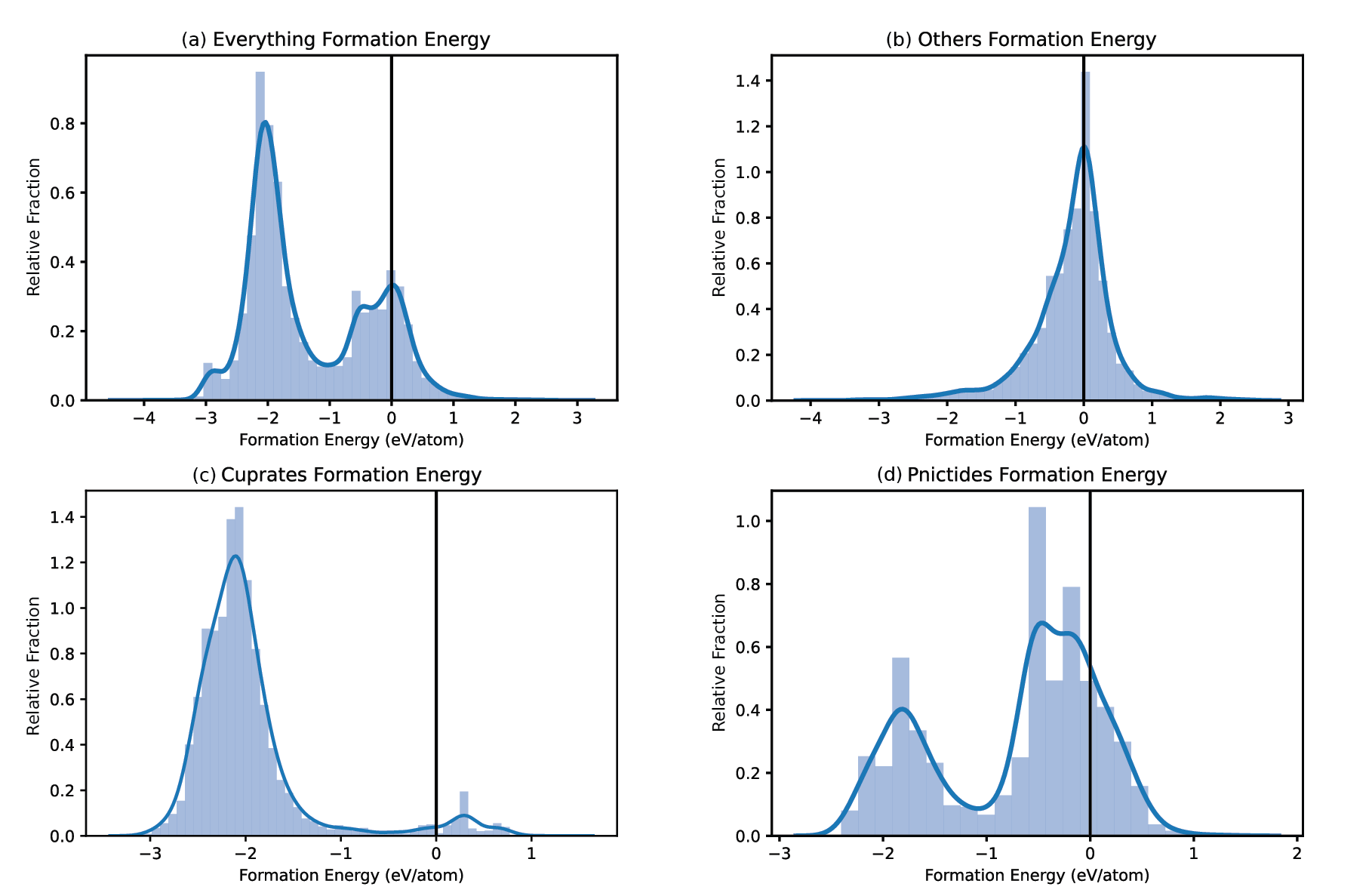}
    \caption{Distribution of the formation energies of the predicted compounds from the four versions of the GAN: (a) everything, (b) others, (c) cuprates, (d) pnictides.}
    \label{fig:form_energy}
\end{figure*}

As mentioned in the previous section, we found the formation energies of the predicted compounds from the GANs using ElemNet \cite{Jha2018}.
It was indicated in Ref.~\cite{Jha2018} that a negative value of formation energy is a good indicator of stability, i.e. the possibility of being 
synthesized in the lab. In Fig.~\ref{fig:form_energy} we display the values of formation energy for all predictions. We see from the 
distributions of formation energies that most of predicted compounds have calculated formation energies less than zero. Even though this does not provide definitive proof of stability, it is a general indication that most of the predicted compounds are stable and 
can be synthesized in the lab.

\subsection{Superconductivity}

As a next test, we ran the predicted compounds through the $K$-Nearest Neighbors (KNN) classification model from Ref.~\cite{ROTER20201353689} for predicting superconductivity based on elemental composition, in order to to check if the predictions of two machine learning models (GAN and KNN) would agree. However, the probabilistic nature of the machine learning model had to be taken into account. If $p_{sc}$ is the proportion of predictions that came up superconducting according to the model, then we can estimate $\rho_{sc}$, the true proportion of superconducting entries, using Bayesian statistics. Denoting $\textit{\textsf{tp}}$ and $\textit{\textsf{fp}}$ as the true positive and false positive rates of the classification model, respectively, we can write
\begin{equation}
    \rho_{sc} \cdot \textit{\textsf{tp}} + (1-\rho) \cdot \textit{\textsf{fp}} \approx p_{sc}.
\end{equation}
Solving for $\rho_{sc}$ gives
\begin{equation}\label{eq:estimate}
    \rho_{sc} \approx \frac{p_{sc} - \textit{\textsf{fp}}}{\textit{\textsf{tp}} - \textit{\textsf{fp}}}.
\end{equation}
The true positive and false positive rates here are reported from Ref.~\cite{ROTER20201353689}:  $\textit{\textsf{tp}} = 98.69\%$ and $\textit{\textsf{fp}} = 16.94\%$. The output percentages along with the estimates of the true proportions calculated from equation \ref{eq:estimate} are tabulated in Table~\ref{tab:supercon_percentages}.

\begin{table}
    \centering
    \begin{ruledtabular}
    \begin{tabular}{ccc}
    GAN Version & Output \% & True \% Estimate \\
    \hline
    Entirety of SuperCon & 74.50 \% & 70.42\%  \\
    Cuprates  & 75.76\% & 71.95\% \\
    Pnictides & 72.44\% & 67.89\% \\
    Others & 69.58\% & 64.39\% \\
    \end{tabular}
    \end{ruledtabular}
    \caption{The percentages of generated predictions that were determined by the KNN model to be superconducting for different training sets along with the estimated real percentage of the predictions that were superconducting. The true percentages were estimated according to Eq.~\ref{eq:estimate}.}
    \label{tab:supercon_percentages}
\end{table}
All four GANs achieved very high percentages of superconducting material according to the KNN model, especially when compared to the 3\% figure from the manual search in Ref.~\cite{Hosono2015}. However, the only definite test of superconductivity can come from experimental measurements.

\subsection{Critical Temperature Estimates}

\begin{figure*}
    \centering
    \includegraphics[scale=0.8]{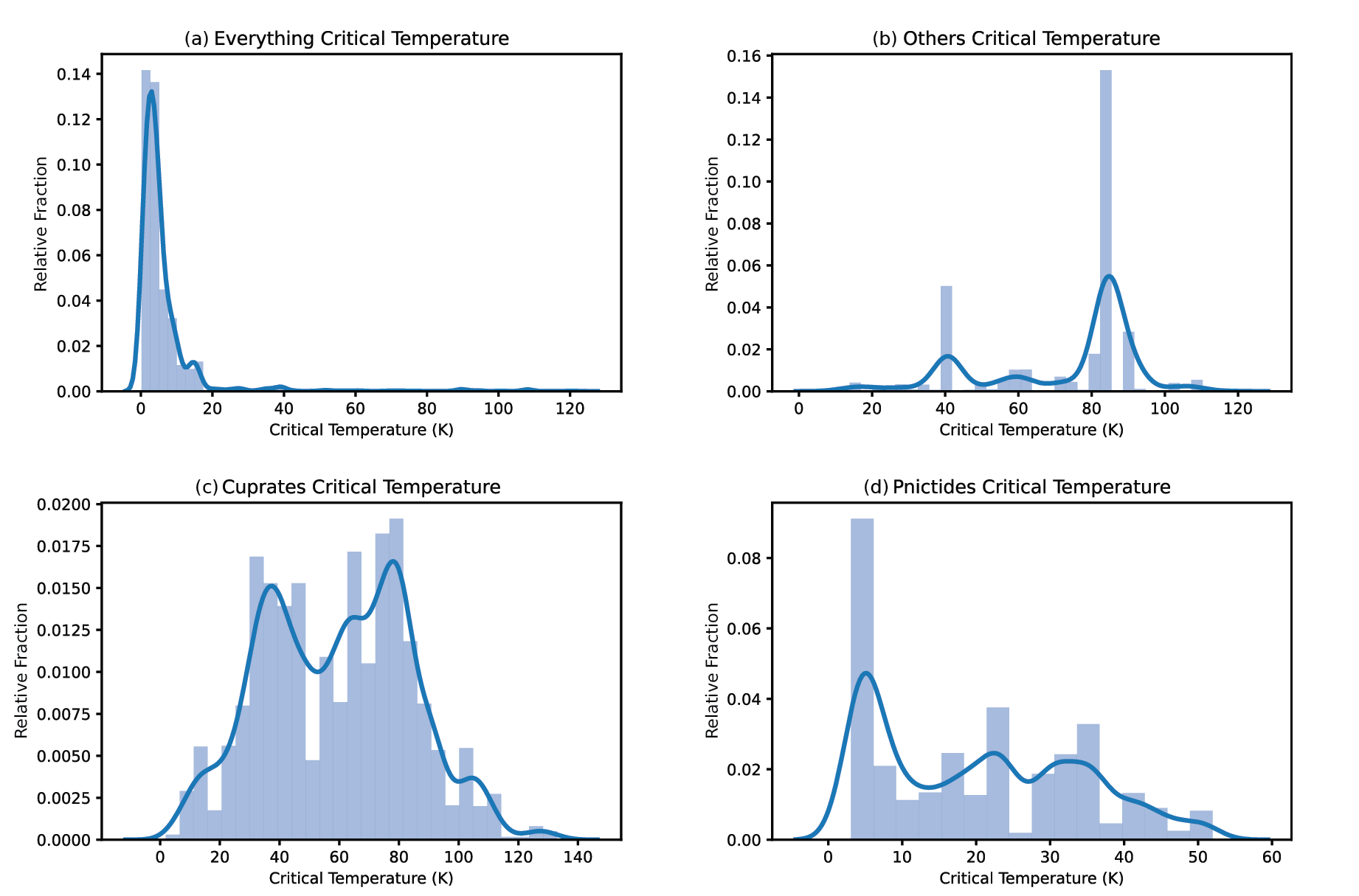}
    \caption{Distributions of the critical temperatures of the predictions of the four different versions of ScGAN: trained on (a) everything, (b) others, (c) cuprates, (d) pnictides. }
    \label{fig:tc_dists}
\end{figure*}

We also calculated the critical temperatures of our predictions using the regression model from Ref.~\cite{ROTER20201353689}. Similar to the superconductivity tests in the previous subsection, these calculated values can only be taken as approximations. However, the regression model can still provide us with a general understanding of the capabilities of ScGAN. The critical temperature outputs from the model are summarized in Table \ref{tab:critical_temps} and the distributions are in shown in Fig.~\ref{fig:tc_dists}. While the distributions are somewhat broad, the GANs were still able to find several superconductors with predicted critical temperatures higher than $100 \, \mathrm{K}$, which exceeds the manual search maximum of $58\, \mathrm{K}$ (though that search was mostly restricted to pnictides) and several of the previous machine learning approaches \cite{ROTER20201353689,Stanev2018}. This is not surprising, as those previous searches were limited to existing databases of stoichiometric compounds, which meant that these forward design approaches could only produce a limited number of candidates, mostly with low critical temperatures.

\begin{table}[t]
    \centering
    \begin{ruledtabular}
    \begin{tabular}{cccc}
    Training Data & Average $T_c$ & Standard Dev. & Max $T_c$ \\
    \hline
    Entirety of SuperCon & $6.53 \, \mathrm{K}$ & $11.76\,\mathrm{K}$ & $123.25\,\mathrm K$  \\
    Cuprates  & $59.34 \, \mathrm{K}$ & $24.78\,\mathrm{K}$ & $133 \,\mathrm K$ \\
    Pnictides & $20.41 \, \mathrm K$ & $13.69 \, \mathrm K$ & $51.98 \, \mathrm K$ \\
    Others & $72.68 \, \mathrm{K}$ & $21.24 \,\mathrm{K}$ & $116.55\,\mathrm K$ \\
    \end{tabular}
    \end{ruledtabular}
    \caption{Summary statistics of the predicted critical temperatures of the generated predictions that were determined by the regression model for different training sets.}
    \label{tab:critical_temps}
\end{table}

\subsection{Ability to Learn Features}

We then looked at the types of superconductors that were generated by the different versions of ScGAN. The distributions are given in Table \ref{tab:by_class}, and we can see that each versions of ScGAN generated mostly superconductors that matched their training data. This indicates the GAN was able to detect the different underlying features behind these different major classes of superconductors.

\begin{table}
    \centering
    \begin{ruledtabular}
    \begin{tabular}{cccc}
        Training Data & Cuprate \% & Pnictide \% & Other \%\\
        \hline
        Cuprate  & 92.76\% &  0.06\% & 7.18\% \\
        Pnictides  & 0.02 \% & 99.84 \% & 0.14\% \\
        Others &  0.14\% & 0.6 \% & 99.26\%
    \end{tabular}
    \end{ruledtabular}
    
    \caption{The distribution of the predicted superconductors across the different classes of superconductors for the different versions of the GAN.}
    \label{tab:by_class}
\end{table}

\subsection{Clustering results}

\begin{figure*}
    \centering
\vspace{-2cm}
\includegraphics[scale=0.5]{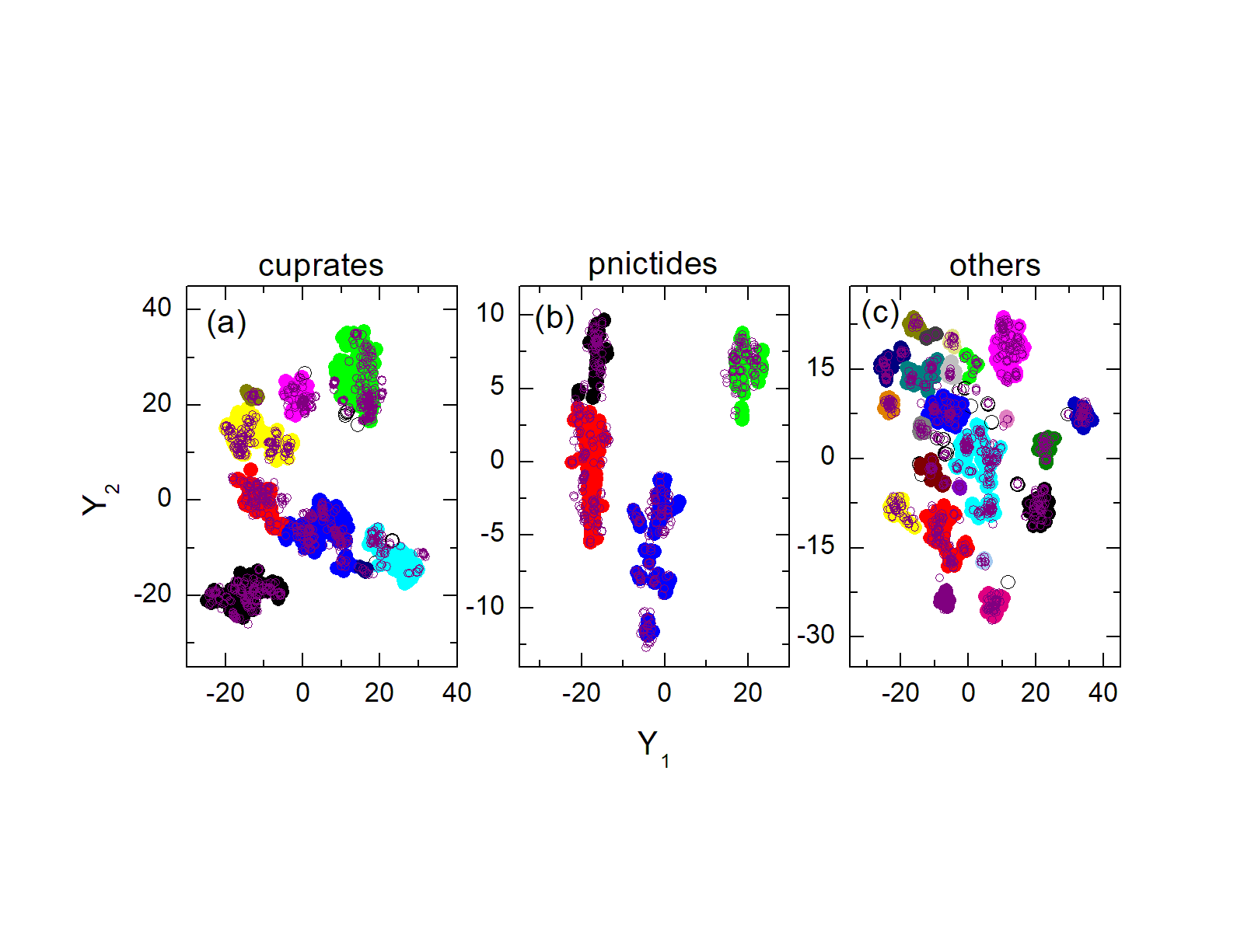}
\vspace{-2cm}
    \caption{Clustering of the predicted compounds 
from various versions of the GAN: (a) cuprates, (b) pnictides and (c) others.
Full circles represent the data points from SuperCon and purple open circles are GAN
predictions.}
    \label{fig:tsne}
\end{figure*}

In Fig.~\ref{fig:tsne} we display the results of 
clustering analysis on three sets of predictions: panel (a)
for cuprates, panel (b) for pnictides and panel (c) for 
others. The results are visualized with the help of t-SNE.
As pointed out above, the two t-SNE dimensions, Y$_1$ and Y$_2$, do not have any physical
meaning. Full circles of different colors represent different families of superconductors
from SuperCon database, whereas purple open circles represent GAN 
predictions. As can be seen from all three panels, GANs
were able to generate new superconductors from all known families 
of cuprates, pnictides and other superconductors. However,
GANs did not predict any new {\it families} of 
superconductors.

\subsection{Promising Candidates}

After running through the $T_c$ prediction model, we manually identified the ones that looked the most promising, including some with very high critical temperatures. It turns out that most of these were cuprates, which is not surprising, considering that superconductors with highest critical temperatures in the training set are cuprates. We then checked the Crystallography Open Database (COD) \cite{Vaitkus2021} to see if those compounds were in the database. The ones listed in Table~\ref{tab:predictions} could not be found neither in COD nor in SuperCon, showing that this model overcomes the limitations of the previous forward design models and finds completely new superconductors. A more comprehensive list of predictions is available upon reasonable request.

\begin{table}
    \centering
    \begin{ruledtabular}
    \begin{tabular}{lcc}
        Compound & Predicted $T_c$ & Class \\ \hline
$\mathrm{PrCaBiSr_{2}Cu_{2}O_{7.46}}$ & $104.6  \, \mathrm K$ & Cuprates \\
$\mathrm{YTiSr_{2}Cu_{2.74}O_{6.76}}$ & $91.7  \, \mathrm K$ & Cuprates \\
$\mathrm{TeYSr_{2}Cu_{2}O_{7.75}}$ & $89.8  \, \mathrm K$ & Cuprates \\
$\mathrm{TlCaSr_{2}Cu_{2}O_{7.82}}$ & $73.9  \, \mathrm K$ & Cuprates \\
$\mathrm{YCaBa_{2}ZnCu_{2.36}O_{7.54}}$ & $71.5  \, \mathrm K$ & Cuprates \\
$\mathrm{HgCsSrCa_{2}Cu_{2.56}O_{8.66}}$ & $69.8  \, \mathrm K$ & Cuprates \\
$\mathrm{GdCaRuSr_{1.83}Cu_{2}O_{8.71}}$ & $40.8  \, \mathrm K$ & Cuprates \\
$\mathrm{C_{2.52}Ni_{0.92}Y_{0.71}Th}$ & $85.3  \, \mathrm K$ & Others \\
$\mathrm{Si_{0.62}V_{0.91}Zr_{0.83}}$ & $84.7  \, \mathrm K$ & Others \\
$\mathrm{Al_{2.34}Te_{0.64}Ir_{1.07}}$ & $84.7  \, \mathrm K$ & Others \\
$\mathrm{Be_{0.16}Si_{1.09}V_{2.67}Y_{1.72}}$ & $62.4  \, \mathrm K$ & Others \\
$\mathrm{Cu_{1.13}Nb_{3.0}Sb_{0.72}Ir_{1.05}}$ & $59.4  \, \mathrm K$ & Others \\
$\mathrm{Ga_{0.62}Nb_{2.88}Sn_{0.65}Te_{0.79}}$ & $40.8  \, \mathrm K$ & Others \\
$\mathrm{B_{1.73}C_{1.03}Ni_{1.12}Y_{0.66}Pt_{0.64}}$ & $40.8  \, \mathrm K$ & Others \\
$\mathrm{RuTeSeFe}$ & $35.6  \, \mathrm K$ & Pnictides \\
$\mathrm{TeSSeFe_{1.05}}$ & $31.  0\, \mathrm K$ & Pnictides \\
$\mathrm{CeCoAs_{2.15}Fe_{1.39}}$ & $23.3  \, \mathrm K$ & Pnictides \\
$\mathrm{CeThPAsFe_{1.59}}$ & $12.2  \, \mathrm K$ & Pnictides \\
$\mathrm{GaPrCa_{2.58}As_{12.44}Fe_{6.34}}$ & $11.9  \, \mathrm K$ & Pnictides \\
$\mathrm{NdOAsFe}$ & $4.5  \, \mathrm K$ & Pnictides \\

    \end{tabular}
    \end{ruledtabular}
    \caption{Promising superconductor candidates generated by ScGANs, that do not exist in current databases. Also shown are their predicted critical temperatures \cite{ROTER20201353689}.}
    \label{tab:predictions}
\end{table}

\section{Conclusion}
\label{conclusions}

For decades the search for new superconductors has relied on the serendipity of material scientists to synthesize a new material with superconducting proprieties. This paper introduced a novel method to search for superconductors---discovering candidates with a generative adversarial network. In contrast to previous computational methods which attempted to predict new superconductors off of existing datasets, this model predicted compounds directly. This ``inverse design'' approach proved to be far more powerful than manual search methods and previous computational methods, with the model being able to generate thousands of candidates with a wide-range of critical temperatures that lied outside of existing databases (both superconductor and general inorganic compound databases). Even while only training on chemical compositions, more than $70\%$ of the GANs predictions were cross-checked with a separate model to be potentially superconducting (the only way to know for sure, however, would be to synthesize and check these compounds in a lab). Of these, several were promising HTS candidates listed in Table \ref{tab:predictions}. We point out that previous models would have been unable to find such candidates as they were outside of current databases.

While the compounds generated were new, our clustering showed that the GAN did not generate any new families of superconductors. However, it was still able to generate non-stoichiometric compounds, widening the scope of the computational search. Future studies should look into some improvements that can be made, such as being able to account for charge neutrality and crystal structure in the compound encodings. Chemical checks on the predicted compounds (using SMACT as detailed earlier) revealed that while there was electronegativity balance, charge balance was not always exact for the superconductor candidates. Furthermore, crystal structure has been known to play a significant role in superconductivity, so including it in calculations would most likely result in improvements. However, this may prove to be a difficult endeavor as crystal structure is not well-documented in existing databases. Active transfer learning could also be attempted to narrow down the predictions to high temperature superconductors only \cite{Kim2021}. However, even without these possible improvements, the model at its current version is still very promising and can be applied to search for superconductors, starting with the candidates identified in this paper.

\bibliography{bibliography}

\providecommand{\noopsort}[1]{}\providecommand{\singleletter}[1]{#1}%
\begin{thebibliography}{23}%
\makeatletter
\providecommand \@ifxundefined [1]{%
 \@ifx{#1\undefined}
}%
\providecommand \@ifnum [1]{%
 \ifnum #1\expandafter \@firstoftwo
 \else \expandafter \@secondoftwo
 \fi
}%
\providecommand \@ifx [1]{%
 \ifx #1\expandafter \@firstoftwo
 \else \expandafter \@secondoftwo
 \fi
}%
\providecommand \natexlab [1]{#1}%
\providecommand \enquote  [1]{``#1''}%
\providecommand \bibnamefont  [1]{#1}%
\providecommand \bibfnamefont [1]{#1}%
\providecommand \citenamefont [1]{#1}%
\providecommand \href@noop [0]{\@secondoftwo}%
\providecommand \href [0]{\begingroup \@sanitize@url \@href}%
\providecommand \@href[1]{\@@startlink{#1}\@@href}%
\providecommand \@@href[1]{\endgroup#1\@@endlink}%
\providecommand \@sanitize@url [0]{\catcode `\\12\catcode `\$12\catcode
  `\&12\catcode `\#12\catcode `\^12\catcode `\_12\catcode `\%12\relax}%
\providecommand \@@startlink[1]{}%
\providecommand \@@endlink[0]{}%
\providecommand \url  [0]{\begingroup\@sanitize@url \@url }%
\providecommand \@url [1]{\endgroup\@href {#1}{\urlprefix }}%
\providecommand \urlprefix  [0]{URL }%
\providecommand \Eprint [0]{\href }%
\providecommand \doibase [0]{https://doi.org/}%
\providecommand \selectlanguage [0]{\@gobble}%
\providecommand \bibinfo  [0]{\@secondoftwo}%
\providecommand \bibfield  [0]{\@secondoftwo}%
\providecommand \translation [1]{[#1]}%
\providecommand \BibitemOpen [0]{}%
\providecommand \bibitemStop [0]{}%
\providecommand \bibitemNoStop [0]{.\EOS\space}%
\providecommand \EOS [0]{\spacefactor3000\relax}%
\providecommand \BibitemShut  [1]{\csname bibitem#1\endcsname}%
\let\auto@bib@innerbib\@empty
%</preamble>
\bibitem [{\citenamefont {Arute}\ \emph {et~al.}(2019)\citenamefont {Arute},
  \citenamefont {Arya}, \citenamefont {Babbush}, \citenamefont {Bacon},
  \citenamefont {Bardin}, \citenamefont {Barends}, \citenamefont {Biswas},
  \citenamefont {Boixo}, \citenamefont {Brandao}, \citenamefont {Buell},
  \citenamefont {Burkett}, \citenamefont {Chen}, \citenamefont {Chen},
  \citenamefont {Chiaro}, \citenamefont {Collins}, \citenamefont {Courtney},
  \citenamefont {Dunsworth}, \citenamefont {Farhi}, \citenamefont {Foxen},\
  and\ \citenamefont {Martinis}}]{Arute2019}%
  \BibitemOpen
  \bibfield  {author} {\bibinfo {author} {\bibfnamefont {F.}~\bibnamefont
  {Arute}}, \bibinfo {author} {\bibfnamefont {K.}~\bibnamefont {Arya}},
  \bibinfo {author} {\bibfnamefont {R.}~\bibnamefont {Babbush}}, \bibinfo
  {author} {\bibfnamefont {D.}~\bibnamefont {Bacon}}, \bibinfo {author}
  {\bibfnamefont {J.~C.}\ \bibnamefont {Bardin}}, \bibinfo {author}
  {\bibfnamefont {R.}~\bibnamefont {Barends}}, \bibinfo {author} {\bibfnamefont
  {R.}~\bibnamefont {Biswas}}, \bibinfo {author} {\bibfnamefont
  {S.}~\bibnamefont {Boixo}}, \bibinfo {author} {\bibfnamefont {F.~G. S.~L.}\
  \bibnamefont {Brandao}}, \bibinfo {author} {\bibfnamefont {D.~A.}\
  \bibnamefont {Buell}}, \bibinfo {author} {\bibfnamefont {B.}~\bibnamefont
  {Burkett}}, \bibinfo {author} {\bibfnamefont {Y.}~\bibnamefont {Chen}},
  \bibinfo {author} {\bibfnamefont {Z.}~\bibnamefont {Chen}}, \bibinfo {author}
  {\bibfnamefont {B.}~\bibnamefont {Chiaro}}, \bibinfo {author} {\bibfnamefont
  {R.}~\bibnamefont {Collins}}, \bibinfo {author} {\bibfnamefont
  {W.}~\bibnamefont {Courtney}}, \bibinfo {author} {\bibfnamefont
  {A.}~\bibnamefont {Dunsworth}}, \bibinfo {author} {\bibfnamefont
  {E.}~\bibnamefont {Farhi}}, \bibinfo {author} {\bibfnamefont
  {B.}~\bibnamefont {Foxen}},\ and\ \bibinfo {author} {\bibfnamefont {J.~M.}\
  \bibnamefont {Martinis}},\ }\bibfield  {title} {\bibinfo {title} {Quantum
  supremacy using a programmable superconducting processor},\ }\href
  {https://doi.org/10.1038/s41586-019-1666-5} {\bibfield  {journal} {\bibinfo
  {journal} {Nature}\ }\textbf {\bibinfo {volume} {574}},\ \bibinfo {pages}
  {505} (\bibinfo {year} {2019})}\BibitemShut {NoStop}%
\bibitem [{\citenamefont {Hirsch}\ \emph {et~al.}(2015)\citenamefont {Hirsch},
  \citenamefont {Maple},\ and\ \citenamefont {Marsiglio}}]{HIRSCH20151}%
  \BibitemOpen
  \bibfield  {author} {\bibinfo {author} {\bibfnamefont {J.}~\bibnamefont
  {Hirsch}}, \bibinfo {author} {\bibfnamefont {M.}~\bibnamefont {Maple}},\ and\
  \bibinfo {author} {\bibfnamefont {F.}~\bibnamefont {Marsiglio}},\ }\bibfield
  {title} {\bibinfo {title} {Superconducting materials classes: Introduction
  and overview},\ }\href
  {https://doi.org/https://doi.org/10.1016/j.physc.2015.03.002} {\bibfield
  {journal} {\bibinfo  {journal} {Physica C: Superconductivity and its
  Applications}\ }\textbf {\bibinfo {volume} {514}},\ \bibinfo {pages} {1}
  (\bibinfo {year} {2015})},\ \bibinfo {note} {superconducting Materials:
  Conventional, Unconventional and Undetermined}\BibitemShut {NoStop}%
\bibitem [{\citenamefont {Hosono}\ \emph {et~al.}(2015)\citenamefont {Hosono},
  \citenamefont {Tanabe}, \citenamefont {Takayama-Muromachi}, \citenamefont
  {Kageyama}, \citenamefont {Yamanaka}, \citenamefont {Kumakura}, \citenamefont
  {Nohara}, \citenamefont {Hiramatsu},\ and\ \citenamefont
  {Fujitsu}}]{Hosono2015}%
  \BibitemOpen
  \bibfield  {author} {\bibinfo {author} {\bibfnamefont {H.}~\bibnamefont
  {Hosono}}, \bibinfo {author} {\bibfnamefont {K.}~\bibnamefont {Tanabe}},
  \bibinfo {author} {\bibfnamefont {E.}~\bibnamefont {Takayama-Muromachi}},
  \bibinfo {author} {\bibfnamefont {H.}~\bibnamefont {Kageyama}}, \bibinfo
  {author} {\bibfnamefont {S.}~\bibnamefont {Yamanaka}}, \bibinfo {author}
  {\bibfnamefont {H.}~\bibnamefont {Kumakura}}, \bibinfo {author}
  {\bibfnamefont {M.}~\bibnamefont {Nohara}}, \bibinfo {author} {\bibfnamefont
  {H.}~\bibnamefont {Hiramatsu}},\ and\ \bibinfo {author} {\bibfnamefont
  {S.}~\bibnamefont {Fujitsu}},\ }\bibfield  {title} {\bibinfo {title}
  {Exploration of new superconductors and functional materials, and fabrication
  of superconducting tapes and wires of iron pnictides},\ }\href
  {https://doi.org/10.1088/1468-6996/16/3/033503} {\bibfield  {journal}
  {\bibinfo  {journal} {Science and Technology of Advanced Materials}\ }\textbf
  {\bibinfo {volume} {16}},\ \bibinfo {pages} {033503} (\bibinfo {year}
  {2015})}\BibitemShut {NoStop}%
\bibitem [{\citenamefont {Bedolla}\ \emph {et~al.}(2020)\citenamefont
  {Bedolla}, \citenamefont {Padierna},\ and\ \citenamefont
  {Casta{\~{n}}eda-Priego}}]{Bedolla_2020}%
  \BibitemOpen
  \bibfield  {author} {\bibinfo {author} {\bibfnamefont {E.}~\bibnamefont
  {Bedolla}}, \bibinfo {author} {\bibfnamefont {L.~C.}\ \bibnamefont
  {Padierna}},\ and\ \bibinfo {author} {\bibfnamefont {R.}~\bibnamefont
  {Casta{\~{n}}eda-Priego}},\ }\bibfield  {title} {\bibinfo {title} {Machine
  learning for condensed matter physics},\ }\href
  {https://doi.org/10.1088/1361-648x/abb895} {\bibfield  {journal} {\bibinfo
  {journal} {Journal of Physics: Condensed Matter}\ }\textbf {\bibinfo {volume}
  {33}},\ \bibinfo {pages} {053001} (\bibinfo {year} {2020})}\BibitemShut
  {NoStop}%
\bibitem [{\citenamefont {Stanev}\ \emph {et~al.}(2018)\citenamefont {Stanev},
  \citenamefont {Oses}, \citenamefont {Kusne}, \citenamefont {Rodriguez},
  \citenamefont {Paglione}, \citenamefont {Curtarolo},\ and\ \citenamefont
  {Takeuchi}}]{Stanev2018}%
  \BibitemOpen
  \bibfield  {author} {\bibinfo {author} {\bibfnamefont {V.}~\bibnamefont
  {Stanev}}, \bibinfo {author} {\bibfnamefont {C.}~\bibnamefont {Oses}},
  \bibinfo {author} {\bibfnamefont {A.~G.}\ \bibnamefont {Kusne}}, \bibinfo
  {author} {\bibfnamefont {E.}~\bibnamefont {Rodriguez}}, \bibinfo {author}
  {\bibfnamefont {J.}~\bibnamefont {Paglione}}, \bibinfo {author}
  {\bibfnamefont {S.}~\bibnamefont {Curtarolo}},\ and\ \bibinfo {author}
  {\bibfnamefont {I.}~\bibnamefont {Takeuchi}},\ }\bibfield  {title} {\bibinfo
  {title} {Machine learning modeling of superconducting critical temperature},\
  }\href {https://doi.org/10.1038/s41524-018-0085-8} {\bibfield  {journal}
  {\bibinfo  {journal} {npj Computational Materials}\ }\textbf {\bibinfo
  {volume} {4}},\ \bibinfo {pages} {29} (\bibinfo {year} {2018})}\BibitemShut
  {NoStop}%
\bibitem [{\citenamefont {Roter}\ and\ \citenamefont
  {Dordevic}(2020)}]{ROTER20201353689}%
  \BibitemOpen
  \bibfield  {author} {\bibinfo {author} {\bibfnamefont {B.}~\bibnamefont
  {Roter}}\ and\ \bibinfo {author} {\bibfnamefont {S.}~\bibnamefont
  {Dordevic}},\ }\bibfield  {title} {\bibinfo {title} {Predicting new
  superconductors and their critical temperatures using machine learning},\
  }\href {https://doi.org/doi.org/10.1016/j.physc.2020.1353689} {\bibfield
  {journal} {\bibinfo  {journal} {Physica C: Superconductivity and its
  Applications}\ }\textbf {\bibinfo {volume} {575}},\ \bibinfo {pages}
  {1353689} (\bibinfo {year} {2020})}\BibitemShut {NoStop}%
\bibitem [{\citenamefont {Konno}\ \emph {et~al.}(2021)\citenamefont {Konno},
  \citenamefont {Kurokawa}, \citenamefont {Nabeshima}, \citenamefont
  {Sakishita}, \citenamefont {Ogawa}, \citenamefont {Hosako},\ and\
  \citenamefont {Maeda}}]{PhysRevB.103.014509}%
  \BibitemOpen
  \bibfield  {author} {\bibinfo {author} {\bibfnamefont {T.}~\bibnamefont
  {Konno}}, \bibinfo {author} {\bibfnamefont {H.}~\bibnamefont {Kurokawa}},
  \bibinfo {author} {\bibfnamefont {F.}~\bibnamefont {Nabeshima}}, \bibinfo
  {author} {\bibfnamefont {Y.}~\bibnamefont {Sakishita}}, \bibinfo {author}
  {\bibfnamefont {R.}~\bibnamefont {Ogawa}}, \bibinfo {author} {\bibfnamefont
  {I.}~\bibnamefont {Hosako}},\ and\ \bibinfo {author} {\bibfnamefont
  {A.}~\bibnamefont {Maeda}},\ }\bibfield  {title} {\bibinfo {title} {Deep
  learning model for finding new superconductors},\ }\href
  {https://doi.org/10.1103/PhysRevB.103.014509} {\bibfield  {journal} {\bibinfo
   {journal} {Phys. Rev. B}\ }\textbf {\bibinfo {volume} {103}},\ \bibinfo
  {pages} {014509} (\bibinfo {year} {2021})}\BibitemShut {NoStop}%
\bibitem [{\citenamefont {Goodfellow}\ \emph {et~al.}(2014)\citenamefont
  {Goodfellow}, \citenamefont {Pouget-Abadie}, \citenamefont {Mirza},
  \citenamefont {Xu}, \citenamefont {Warde-Farley}, \citenamefont {Ozair},
  \citenamefont {Courville},\ and\ \citenamefont {Bengio}}]{goodfellow2014}%
  \BibitemOpen
  \bibfield  {author} {\bibinfo {author} {\bibfnamefont {I.~J.}\ \bibnamefont
  {Goodfellow}}, \bibinfo {author} {\bibfnamefont {J.}~\bibnamefont
  {Pouget-Abadie}}, \bibinfo {author} {\bibfnamefont {M.}~\bibnamefont
  {Mirza}}, \bibinfo {author} {\bibfnamefont {B.}~\bibnamefont {Xu}}, \bibinfo
  {author} {\bibfnamefont {D.}~\bibnamefont {Warde-Farley}}, \bibinfo {author}
  {\bibfnamefont {S.}~\bibnamefont {Ozair}}, \bibinfo {author} {\bibfnamefont
  {A.}~\bibnamefont {Courville}},\ and\ \bibinfo {author} {\bibfnamefont
  {Y.}~\bibnamefont {Bengio}},\ }\bibfield  {title} {\bibinfo {title}
  {Generative adversarial nets},\ }in\ \href@noop {} {\emph {\bibinfo
  {booktitle} {Proceedings of the 27th International Conference on Neural
  Information Processing Systems - Volume 2}}},\ \bibinfo {series and number}
  {NIPS'14}\ (\bibinfo  {publisher} {MIT Press},\ \bibinfo {address}
  {Cambridge, MA, USA},\ \bibinfo {year} {2014})\ p.\ \bibinfo {pages}
  {2672–2680}\BibitemShut {NoStop}%
\bibitem [{\citenamefont {Dash}\ \emph {et~al.}(2021)\citenamefont {Dash},
  \citenamefont {Ye},\ and\ \citenamefont {Wang}}]{DBLP:ganreview}%
  \BibitemOpen
  \bibfield  {author} {\bibinfo {author} {\bibfnamefont {A.}~\bibnamefont
  {Dash}}, \bibinfo {author} {\bibfnamefont {J.}~\bibnamefont {Ye}},\ and\
  \bibinfo {author} {\bibfnamefont {G.}~\bibnamefont {Wang}},\ }\bibfield
  {title} {\bibinfo {title} {A review of generative adversarial networks (gans)
  and its applications in a wide variety of disciplines--from medical to remote
  sensing},\ }\href {https://arxiv.org/abs/2110.01442} {\bibfield  {journal}
  {\bibinfo  {journal} {arXiv preprint arXiv:2110.01442}\ } (\bibinfo {year}
  {2021})}\BibitemShut {NoStop}%
\bibitem [{\citenamefont {Dan}\ \emph {et~al.}(2020)\citenamefont {Dan},
  \citenamefont {Zhao}, \citenamefont {Li}, \citenamefont {Li}, \citenamefont
  {Hu},\ and\ \citenamefont {Hu}}]{Dan2020}%
  \BibitemOpen
  \bibfield  {author} {\bibinfo {author} {\bibfnamefont {Y.}~\bibnamefont
  {Dan}}, \bibinfo {author} {\bibfnamefont {Y.}~\bibnamefont {Zhao}}, \bibinfo
  {author} {\bibfnamefont {X.}~\bibnamefont {Li}}, \bibinfo {author}
  {\bibfnamefont {S.}~\bibnamefont {Li}}, \bibinfo {author} {\bibfnamefont
  {M.}~\bibnamefont {Hu}},\ and\ \bibinfo {author} {\bibfnamefont
  {J.}~\bibnamefont {Hu}},\ }\bibfield  {title} {\bibinfo {title} {Generative
  adversarial networks (gan) based efficient sampling of chemical composition
  space for inverse design of inorganic materials},\ }\href
  {https://doi.org/10.1038/s41524-020-00352-0} {\bibfield  {journal} {\bibinfo
  {journal} {npj Computational Materials}\ }\textbf {\bibinfo {volume} {6}},\
  \bibinfo {pages} {84} (\bibinfo {year} {2020})}\BibitemShut {NoStop}%
\bibitem [{\citenamefont {Hu}\ \emph {et~al.}(2020)\citenamefont {Hu},
  \citenamefont {Song}, \citenamefont {Jiang},\ and\ \citenamefont
  {Li}}]{sym12111889}%
  \BibitemOpen
  \bibfield  {author} {\bibinfo {author} {\bibfnamefont {T.}~\bibnamefont
  {Hu}}, \bibinfo {author} {\bibfnamefont {H.}~\bibnamefont {Song}}, \bibinfo
  {author} {\bibfnamefont {T.}~\bibnamefont {Jiang}},\ and\ \bibinfo {author}
  {\bibfnamefont {S.}~\bibnamefont {Li}},\ }\bibfield  {title} {\bibinfo
  {title} {Learning representations of inorganic materials from generative
  adversarial networks},\ }\bibfield  {journal} {\bibinfo  {journal}
  {Symmetry}\ }\textbf {\bibinfo {volume} {12}},\ \href
  {https://doi.org/10.3390/sym12111889} {10.3390/sym12111889} (\bibinfo {year}
  {2020})\BibitemShut {NoStop}%
\bibitem [{\citenamefont {{National Institute for Materials
  Science}}()}]{supercon.nims.go.jp}%
  \BibitemOpen
  \bibfield  {author} {\bibinfo {author} {\bibnamefont {{National Institute for
  Materials Science}}},\ }\href {https://supercon.nims.go.jp/en/}
  {}\BibitemShut {NoStop}%
\bibitem [{\citenamefont {Saal}\ \emph {et~al.}(2013)\citenamefont {Saal},
  \citenamefont {Kirklin}, \citenamefont {Aykol}, \citenamefont {Meredig},\
  and\ \citenamefont {Wolverton}}]{Saal2013}%
  \BibitemOpen
  \bibfield  {author} {\bibinfo {author} {\bibfnamefont {J.~E.}\ \bibnamefont
  {Saal}}, \bibinfo {author} {\bibfnamefont {S.}~\bibnamefont {Kirklin}},
  \bibinfo {author} {\bibfnamefont {M.}~\bibnamefont {Aykol}}, \bibinfo
  {author} {\bibfnamefont {B.}~\bibnamefont {Meredig}},\ and\ \bibinfo {author}
  {\bibfnamefont {C.}~\bibnamefont {Wolverton}},\ }\bibfield  {title} {\bibinfo
  {title} {Materials design and discovery with high-throughput density
  functional theory: The open quantum materials database (oqmd)},\ }\href
  {https://doi.org/10.1007/s11837-013-0755-4} {\bibfield  {journal} {\bibinfo
  {journal} {JOM}\ }\textbf {\bibinfo {volume} {65}},\ \bibinfo {pages} {1501}
  (\bibinfo {year} {2013})}\BibitemShut {NoStop}%
\bibitem [{\citenamefont {Kirklin}\ \emph {et~al.}(2015)\citenamefont
  {Kirklin}, \citenamefont {Saal}, \citenamefont {Meredig}, \citenamefont
  {Thompson}, \citenamefont {Doak}, \citenamefont {Aykol}, \citenamefont
  {R{\"u}hl},\ and\ \citenamefont {Wolverton}}]{Kirklin2015}%
  \BibitemOpen
  \bibfield  {author} {\bibinfo {author} {\bibfnamefont {S.}~\bibnamefont
  {Kirklin}}, \bibinfo {author} {\bibfnamefont {J.~E.}\ \bibnamefont {Saal}},
  \bibinfo {author} {\bibfnamefont {B.}~\bibnamefont {Meredig}}, \bibinfo
  {author} {\bibfnamefont {A.}~\bibnamefont {Thompson}}, \bibinfo {author}
  {\bibfnamefont {J.~W.}\ \bibnamefont {Doak}}, \bibinfo {author}
  {\bibfnamefont {M.}~\bibnamefont {Aykol}}, \bibinfo {author} {\bibfnamefont
  {S.}~\bibnamefont {R{\"u}hl}},\ and\ \bibinfo {author} {\bibfnamefont
  {C.}~\bibnamefont {Wolverton}},\ }\bibfield  {title} {\bibinfo {title} {The
  open quantum materials database (oqmd): assessing the accuracy of dft
  formation energies},\ }\href {https://doi.org/10.1038/npjcompumats.2015.10}
  {\bibfield  {journal} {\bibinfo  {journal} {npj Computational Materials}\
  }\textbf {\bibinfo {volume} {1}},\ \bibinfo {pages} {15010} (\bibinfo {year}
  {2015})}\BibitemShut {NoStop}%
\bibitem [{\citenamefont {Roter}\ \emph {et~al.}(2022)\citenamefont {Roter},
  \citenamefont {Ninkovic},\ and\ \citenamefont {Dordevic}}]{ROTER20221354078}%
  \BibitemOpen
  \bibfield  {author} {\bibinfo {author} {\bibfnamefont {B.}~\bibnamefont
  {Roter}}, \bibinfo {author} {\bibfnamefont {N.}~\bibnamefont {Ninkovic}},\
  and\ \bibinfo {author} {\bibfnamefont {S.}~\bibnamefont {Dordevic}},\
  }\bibfield  {title} {\bibinfo {title} {Clustering superconductors using
  unsupervised machine learning},\ }\href
  {https://doi.org/https://doi.org/10.1016/j.physc.2022.1354078} {\bibfield
  {journal} {\bibinfo  {journal} {Physica C: Superconductivity and its
  Applications}\ }\textbf {\bibinfo {volume} {598}},\ \bibinfo {pages}
  {1354078} (\bibinfo {year} {2022})}\BibitemShut {NoStop}%
\bibitem [{\citenamefont {Gulrajani}\ \emph {et~al.}(2017)\citenamefont
  {Gulrajani}, \citenamefont {Ahmed}, \citenamefont {Arjovsky}, \citenamefont
  {Dumoulin},\ and\ \citenamefont
  {Courville}}]{DBLP:journals/corr/GulrajaniAADC17}%
  \BibitemOpen
  \bibfield  {author} {\bibinfo {author} {\bibfnamefont {I.}~\bibnamefont
  {Gulrajani}}, \bibinfo {author} {\bibfnamefont {F.}~\bibnamefont {Ahmed}},
  \bibinfo {author} {\bibfnamefont {M.}~\bibnamefont {Arjovsky}}, \bibinfo
  {author} {\bibfnamefont {V.}~\bibnamefont {Dumoulin}},\ and\ \bibinfo
  {author} {\bibfnamefont {A.~C.}\ \bibnamefont {Courville}},\ }\bibfield
  {title} {\bibinfo {title} {Improved training of wasserstein gans},\
  }\href@noop {} {\bibfield  {journal} {\bibinfo  {journal} {Advances in neural
  information processing systems}\ }\textbf {\bibinfo {volume} {30}} (\bibinfo
  {year} {2017})}\BibitemShut {NoStop}%
\bibitem [{\citenamefont {Hinton}(2012)}]{hintonrmsprop}%
  \BibitemOpen
  \bibfield  {author} {\bibinfo {author} {\bibfnamefont {G.}~\bibnamefont
  {Hinton}},\ }\href
  {http://www.cs.toronto.edu/~tijmen/csc321/slides/lecture_slides_lec6.pdf}
  {\bibinfo {title} {Neural networks for machine learning: Lecture 6a overview
  of min-batch gradient descent}} (\bibinfo {year} {2012})\BibitemShut
  {NoStop}%
\bibitem [{\citenamefont {Kingma}\ and\ \citenamefont
  {Ba}(2014)}]{kingma2014adam}%
  \BibitemOpen
  \bibfield  {author} {\bibinfo {author} {\bibfnamefont {D.~P.}\ \bibnamefont
  {Kingma}}\ and\ \bibinfo {author} {\bibfnamefont {J.}~\bibnamefont {Ba}},\
  }\bibfield  {title} {\bibinfo {title} {Adam: A method for stochastic
  optimization},\ }\href@noop {} {\bibfield  {journal} {\bibinfo  {journal}
  {arXiv preprint arXiv:1412.6980}\ } (\bibinfo {year} {2014})}\BibitemShut
  {NoStop}%
\bibitem [{\citenamefont {Davies}\ \emph {et~al.}(2016)\citenamefont {Davies},
  \citenamefont {Butler}, \citenamefont {Jackson}, \citenamefont {Morris},
  \citenamefont {Frost}, \citenamefont {Skelton},\ and\ \citenamefont
  {Walsh}}]{DAVIES2016617}%
  \BibitemOpen
  \bibfield  {author} {\bibinfo {author} {\bibfnamefont {D.}~\bibnamefont
  {Davies}}, \bibinfo {author} {\bibfnamefont {K.}~\bibnamefont {Butler}},
  \bibinfo {author} {\bibfnamefont {A.}~\bibnamefont {Jackson}}, \bibinfo
  {author} {\bibfnamefont {A.}~\bibnamefont {Morris}}, \bibinfo {author}
  {\bibfnamefont {J.}~\bibnamefont {Frost}}, \bibinfo {author} {\bibfnamefont
  {J.}~\bibnamefont {Skelton}},\ and\ \bibinfo {author} {\bibfnamefont
  {A.}~\bibnamefont {Walsh}},\ }\bibfield  {title} {\bibinfo {title}
  {Computational screening of all stoichiometric inorganic materials},\ }\href
  {https://doi.org/10.1016/j.chempr.2016.09.010} {\bibfield  {journal}
  {\bibinfo  {journal} {Chem}\ }\textbf {\bibinfo {volume} {1}},\ \bibinfo
  {pages} {617} (\bibinfo {year} {2016})}\BibitemShut {NoStop}%
\bibitem [{\citenamefont {Jha}\ \emph {et~al.}(2018)\citenamefont {Jha},
  \citenamefont {Ward}, \citenamefont {Paul}, \citenamefont {Liao},
  \citenamefont {Choudhary}, \citenamefont {Wolverton},\ and\ \citenamefont
  {Agrawal}}]{Jha2018}%
  \BibitemOpen
  \bibfield  {author} {\bibinfo {author} {\bibfnamefont {D.}~\bibnamefont
  {Jha}}, \bibinfo {author} {\bibfnamefont {L.}~\bibnamefont {Ward}}, \bibinfo
  {author} {\bibfnamefont {A.}~\bibnamefont {Paul}}, \bibinfo {author}
  {\bibfnamefont {W.-k.}\ \bibnamefont {Liao}}, \bibinfo {author}
  {\bibfnamefont {A.}~\bibnamefont {Choudhary}}, \bibinfo {author}
  {\bibfnamefont {C.}~\bibnamefont {Wolverton}},\ and\ \bibinfo {author}
  {\bibfnamefont {A.}~\bibnamefont {Agrawal}},\ }\bibfield  {title} {\bibinfo
  {title} {Elemnet: Deep learning the chemistry of materials from only
  elemental composition},\ }\href {https://doi.org/10.1038/s41598-018-35934-y}
  {\bibfield  {journal} {\bibinfo  {journal} {Scientific Reports}\ }\textbf
  {\bibinfo {volume} {8}},\ \bibinfo {pages} {17593} (\bibinfo {year}
  {2018})}\BibitemShut {NoStop}%
\bibitem [{\citenamefont {van~der Maaten}\ and\ \citenamefont
  {Hinton}(2008)}]{vanDerMaaten2008}%
  \BibitemOpen
  \bibfield  {author} {\bibinfo {author} {\bibfnamefont {L.}~\bibnamefont
  {van~der Maaten}}\ and\ \bibinfo {author} {\bibfnamefont {G.}~\bibnamefont
  {Hinton}},\ }\bibfield  {title} {\bibinfo {title} {Visualizing data using
  {t-SNE}},\ }\href {http://www.jmlr.org/papers/v9/vandermaaten08a.html}
  {\bibfield  {journal} {\bibinfo  {journal} {Journal of Machine Learning
  Research}\ }\textbf {\bibinfo {volume} {9}},\ \bibinfo {pages} {2579}
  (\bibinfo {year} {2008})}\BibitemShut {NoStop}%
\bibitem [{\citenamefont {Vaitkus}\ \emph {et~al.}(2021)\citenamefont
  {Vaitkus}, \citenamefont {Merkys},\ and\ \citenamefont
  {Gražulis}}]{Vaitkus2021}%
  \BibitemOpen
  \bibfield  {author} {\bibinfo {author} {\bibfnamefont {A.}~\bibnamefont
  {Vaitkus}}, \bibinfo {author} {\bibfnamefont {A.}~\bibnamefont {Merkys}},\
  and\ \bibinfo {author} {\bibfnamefont {S.}~\bibnamefont {Gražulis}},\
  }\bibfield  {title} {\bibinfo {title} {Validation of the {C}rystallography
  {O}pen {D}atabase using the {C}rystallographic {I}nformation {F}ramework},\
  }\href {https://doi.org/10.1107/S1600576720016532} {\bibfield  {journal}
  {\bibinfo  {journal} {Journal of Applied Crystallography}\ }\textbf {\bibinfo
  {volume} {54}},\ \bibinfo {pages} {661} (\bibinfo {year} {2021})}\BibitemShut
  {NoStop}%
\bibitem [{\citenamefont {Kim}\ \emph {et~al.}(2021)\citenamefont {Kim},
  \citenamefont {Kim}, \citenamefont {Yang}, \citenamefont {Park},
  \citenamefont {Gu},\ and\ \citenamefont {Ryu}}]{Kim2021}%
  \BibitemOpen
  \bibfield  {author} {\bibinfo {author} {\bibfnamefont {Y.}~\bibnamefont
  {Kim}}, \bibinfo {author} {\bibfnamefont {Y.}~\bibnamefont {Kim}}, \bibinfo
  {author} {\bibfnamefont {C.}~\bibnamefont {Yang}}, \bibinfo {author}
  {\bibfnamefont {K.}~\bibnamefont {Park}}, \bibinfo {author} {\bibfnamefont
  {G.~X.}\ \bibnamefont {Gu}},\ and\ \bibinfo {author} {\bibfnamefont
  {S.}~\bibnamefont {Ryu}},\ }\bibfield  {title} {\bibinfo {title} {Deep
  learning framework for material design space exploration using active
  transfer learning and data augmentation},\ }\href
  {https://doi.org/10.1038/s41524-021-00609-2} {\bibfield  {journal} {\bibinfo
  {journal} {npj Computational Materials}\ }\textbf {\bibinfo {volume} {7}},\
  \bibinfo {pages} {140} (\bibinfo {year} {2021})}\BibitemShut {NoStop}%
\end{thebibliography}%

\end{document}